\documentstyle[12pt,eqsecnum,aps]{revtex}
\input{epsf.tex}
\begin{document}
\baselineskip=17.pt
\preprint{}
\title{GRAVITON SPECTRA IN STRING COSMOLOGY}
\author{Massimo Galluccio$^1$, Marco Litterio$^2$ and Franco Occhionero$^1$}
\address{$^1$ Osservatorio Astronomico di Roma, Via del Parco Mellini 84, 00136
ROMA--IT\\
$^2$Istituto Astronomico dell'Universit\`a ``La Sapienza'',
 Via G. M. Lancisi 29, 00186 ROMA--IT}
\date{\today}
\maketitle
\vspace{.5truecm}
\begin{abstract}
\baselineskip=17.pt
We propose to uncover the signature of a stringy era 
in the primordial Universe by searching for a prominent
peak in the relic graviton spectrum. This feature, which 
in our specific model
terminates an $\omega^3$ increase and initiates an $\omega^{-7}$
decrease, is induced during the so far overlooked bounce
of the scale factor between the collapsing deflationary
era (or pre-Big Bang) and the expanding inflationary
era (or post-Big Bang).
We evaluate both analytically and numerically the frequency and 
the intensity of the peak and we show that they may likely fall
in the realm of the new generation of interferometric detectors.
The existence of a peak is at variance with ordinarily monotonic
(either increasing or decreasing) graviton spectra of canonical
cosmologies; its detection would therefore offer strong support
to string cosmology.
\end{abstract}
\section{Introduction}
A stochastic background of gravitational waves (GW) is produced during 
inflationary eras 
(whether exponential, power-law or pole-like). The resulting spectra 
differ from each other, depending upon the underlying inflationary model.
Thus one could in principle obtain information about 
the physics at the Planck epoch from the 
observation of the GW background. 

The most appealing candidate for a description of physics at the Planck scale 
is superstring theory: hence comes therefore the most promising 
cosmological model.

The crucial problem is to give a detailed and consistent description of the 
Universe approaching the singularity. By 
using the $O(D-1,D-1)$  simmetry of the tree--level superstring action \cite{odd}
\begin{equation}\label{action}
S=-\frac{1}{2l_{st}^{D-2}}\int\! \mbox d^Dx\sqrt{-g}\mbox e^{-\phi}
\!\left[
R+\nabla_\mu\phi\nabla^\mu\phi
+V(\phi)\right]\,\,,
\end{equation}
where $D\geq 4$, one can  construct  non--singular cosmologies 
\cite{gvpbb}.
The fundamental requirement for such a model is the existence of the so called 
``branch--changing" solutions \cite{bc} which  smoothly interpolate between a 
contracting Universe ({\em pre--big--bang}, PBB) and an expanding one 
({\em post--big--bang}), related  to one another by a duality transformation. 
 This feature is quite general. In fact,
 the full string loop 
expansion, that  includes
the case where the spatial 
hypersurfaces have non--zero curvature, shows ``bouncing" solutions
resembling the ``branch--changing"  ones \cite{loop}. 

In PBB a pole-like inflation
(also referred to as superinflation) is always present at $t<0$ due to the
theory symmetries (foremost among these a particular $O(D-1,D-1)$ symmetry
known as scale factor duality, SFD). In fact, under
SFD the ordinary Friedmann-Robertson-Walker radiation dominated solution
(FRW)  $a(t)\propto t^{1/2}, t>0$ is mapped into the superinflationary
one $a(t)\propto (-t)^{-1/2}, t<0$. The two eras can be smoothly connected
({\em graceful exit})
if a potential $V(\Phi)$ for the dilaton $\Phi$ is 
provided.\footnote{Alternatively the two eras can be connected by a quantum 
cosmological scattering \cite{qustr}.} 
Around $t=0$ the curvature of space--time approaches the string scale, the 
dilaton settles down to a minimum of its potential and the coupling evolves 
towards the present ratio of the Planck to the string scale \cite{bc}
\[
\frac{l_P}{l_{st}}\simeq g_s (\sim g_{\mbox{\scriptsize gauge}}) = 
\mbox e^{\phi/2} \: \rightarrow\:10^{-1}\div 10^{-2}.
\]
After this high curvature regime (stringy era) the Universe reexpands again, 
finally joining the $a(t)\propto t^{1/2}$  radiation dominated era at 
$t\rightarrow \infty$.

This picture holds in the string frame (SF), where the
dilaton is non--minimally coupled to the scalar curvature. However the SF
is not the most convenient for our objectives: for instance, the
exact solution for the dilaton during the joining era is not known. 
On the contrary in the Einstein frame (EF, minimal coupling) 
the dependence from the dilaton of the GW equation disappears,
which is a very good reason for staying henceforth in the EF. 
Here the PBB picture partially changes, in that 
superinflation is turned into accelerated contraction, $H:=\dot a/a<0$, 
$\dot H<0$,  {\em i.e.} into
{\em deflation}. Now, it has been shown \cite{defla} that deflation is, 
under almost all the effects, equivalent to inflation. In particular this
is true for the background gravitational spectrum \cite{bggmv}. We will 
exploit this equivalence whenever deflation appears in cosmological
solutions.
After the conformal transformation to the EF
\[
g_{\mu\nu}^{(EF)}=g_{\mu\nu}^{(SF)}\mbox e^{-\phi^{(SF)}},\quad\quad
\mbox dt^{(EF)}=\mbox e^{-\phi^{(SF)}/2}\mbox dt^{(SF)},
\]
one gets \cite{vacuum} for $t\ll 0$ and $D=4$:
\begin{equation}
 a_{EF}(t_{EF})\propto (-t_{EF})^{{1}/{3}},\quad 
 \phi_{EF}(t_{EF})\propto -\left(\frac{2}{\sqrt 3}\right)\ln (-t_{EF})\,\,.
\end{equation}

Deflation appears as the EF expression of superinflation in the SF,
PBB scenario.
This means that as a general feature the universe contracts, reaches a 
minimum size and after that reexpands. The detailed behaviour of the 
scale factor $a(t)$ depends on the model. It is possible, however, to 
extract the 
qualitative effects from general considerations, independently from the
knowledge of the exact solutions for $a(t)$ and $\phi(t)$. In fact the classical
GW equation  [see (\ref{clwa}) below] does not depend 
explicitly on the dilaton dynamics: in fact,
it is sufficient to fix the evolution of the metric, regardless of that
of the dilaton.
Obviously, the era around the minimum between collapse  and  expansion, that is
commonly referred to as {\it stringy},
the solution can be approximated with:
\begin{equation}\label{eq1b}
a=a_{min}+\left(\frac{t}{t_0}\right)^2\,,\qquad t\simeq 0\,\,. 
\end{equation}

A clear example of complete solution of string cosmology  in the SF
is reported
in \cite{aalo}. It shows the expected three phases: 1) the $t\to -\infty$
inflationary solution; 2) the $t\simeq 0$ smooth connection in the full 
stringy era; 3) the post-Big-Bang radiation dominated era. With respect to
the original PBB model a new feature arises: 
in the full stringy era the scale factor 
$a(t)$ decreases at first, reaches a minimum and reexpands, finally joining the
$t^{1/2}$ run. 
It turns out that the contraction is accelerated, i.e. deflation appears 
again. 

We will exploit the property of string cosmological models of possessing
a deflationary era. In particular, this paper is devoted to studying the
modifications in the spectrum of primordial GW's 
produced by the stringy era joining the deflationary stage to the radiation
dominated one.

This paper is organized as follows. Sec.II is dedicated to the graviton spectra.
We first introduce our formula for the energy density per octave $\Omega_g$.
After that, in Sec. II.A we show the known results for the spectrum from 
the dilaton era; 
in Sec. II.B we present our new results on the contributions from the
stringy era. Our conclusions are reported in Sec. III.

\section{Primordial graviton spectra}
Spectra of primordial gravitational waves are well known.
Differences in the form of the spectra depend on the evolution 
of the cosmological
background in the two relevant phases: the {\em goodbye} phase in which
the wavelength becomes larger than the effective horizon, $H^{-1}$, and the
{\em hallo again} phase in which the waves reenter the horizon. Suppose
$a\propto t^b$ (for $b\to\infty$ exponential inflation is recovered) in 
the former and $a\propto t^c$ in the latter.
For exponential inflation, the fraction of critical density
contributed per octave $\Omega(k)$ does not depend on $k$, i.e. the spectrum
is flat;
in the power-law case the spectrum is decreasing; in the pole-like one
the spectrum increases. The different behaviors can be collected in one
formula:
\begin{equation}\label{eq1}
\Omega(k)\propto k^{\beta}\quad, \qquad 
\beta\equiv2\frac{3c+b-2bc-2}{(1-b)(c-1)}\,\,,
\end{equation}
which can be easily derived with both quantum and classical
arguments and is given in a slightly different (but fully equivalent) form 
by Sahni \cite{allen2}. 
The quantum derivation consists in considering the asymptotic
wave--equation solutions that are 
related by a Bogoliubov transformation \cite{allen2}. 
(The complete solutions are reported 
in \cite{Gasperini} and  in \cite{Buonanno}.)
The positive frequency modes (the asymptotically vacuum state at $t\rightarrow 
-\infty$) will be in general a linear superposition of modes of 
positive and negative frequencies with respect to the vacuum to the right 
$(t\rightarrow +\infty$). The coefficient of the negative mode of the 
 vacuum to the right, determines the number of gravitons created by the 
 interaction 
with the background curvature \cite{allen2}:
\[
\Omega_g^q(\nu):=\frac{1}{\rho_{cr}}\frac{\mbox d \rho_g}{\mbox d\ln \nu}
\propto \frac{\nu^4}{\rho_{cr}}\left|c_-(\nu)\right|^2\,\,.
\]

At the classical level the following expression holds: 
\[
\Omega_g^{c}(\omega):=\frac{1}{\rho_{cr}}\frac{\mbox d\rho_g}{\mbox d\ln \omega}
=\frac{\pi c^2}{4G\rho_{cr}}\omega^2\left|\Delta h_k\right|^2\,\,;
\]
here $\left|\Delta h_k\right|:=k^3h_k^2$ is the spectral amplitude,
$\omega(t)=k/a(t)$ and $h_k(t)$ is the amplitude of the linear wave of 
wave vector $k$. 
One gets\footnote{A rigorous proof of the 
equivalence  of the equation in the different frames is 
reported in \cite{TESI}}  formula (\ref{eq1}) in the EF,
by considering the classical solution of the wave equation 
\begin{equation}\label{clwa}
\ddot h_k+3 H\dot h_k+\omega^2h_k=0\,\,,
\end{equation}
where a dot means derivation wrt standard time.

Note that in the model discussed here new problems arise. In fact, as the 
the scale factor undergoes a bounce, the modes that exit the horizon during  
the dilaton--driven era (PBB), necessarily cross it again twice 
during the pure stringy era (see Sec. II.B and Fig.~\ref{wave} below). 
Furthermore   
their amplitudes grow, instead of being frozen once outside the horizon.
As a consequence the quantum derivation cannot be straightforwardly extended. 
On the contrary, one can still apply the classical derivation:
 with the appropriate 
correction for the bounce in the form of $a(t)\propto (-t)^g$,
$t<0,g>1$, we find the new analytic result \cite{TESI}:
\begin{equation}
\Omega_g\propto k^{\beta'},\quad\quad \beta'=\beta+2(3g-1)/(1-g)\,\,,
\label{neweq}\end{equation}
which is fully supported by numerical computations.
 
\subsection{Spectrum from the dilaton era}
\begin{figure}
\hskip 2truecm
\epsfxsize=11cm
\epsffile{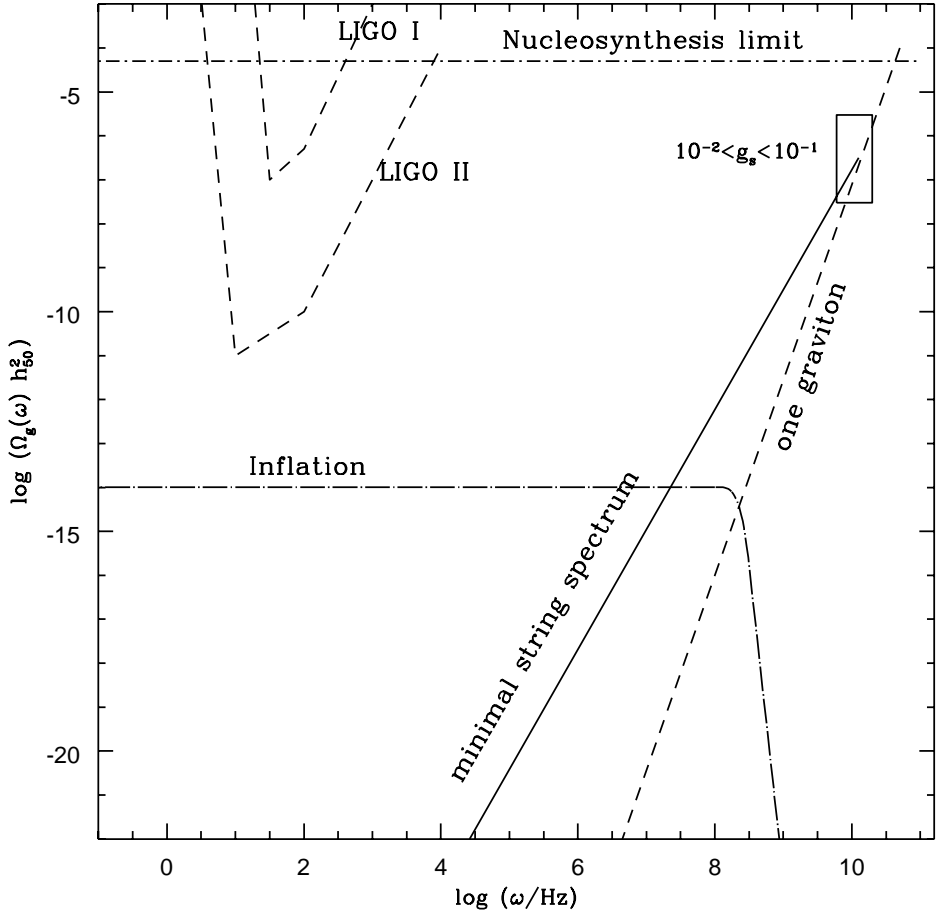}
\caption{\baselineskip 15.pt 
The spectra in the two cases above compared to the LIGO 
 sensitivity, and to the nucleosynthesis bound. Also shown is the 
limit for graviton creation}\label{spettro1}
\end{figure}

{}From (\ref{eq1}) it is easily seen that for deflation+radiation the
spectrum is growing in $k$ FIG.\ref{spettro1}, like in the superinflation 
case. 

\begin{figure}
\hskip 2truecm
\epsfxsize=11cm
\epsffile{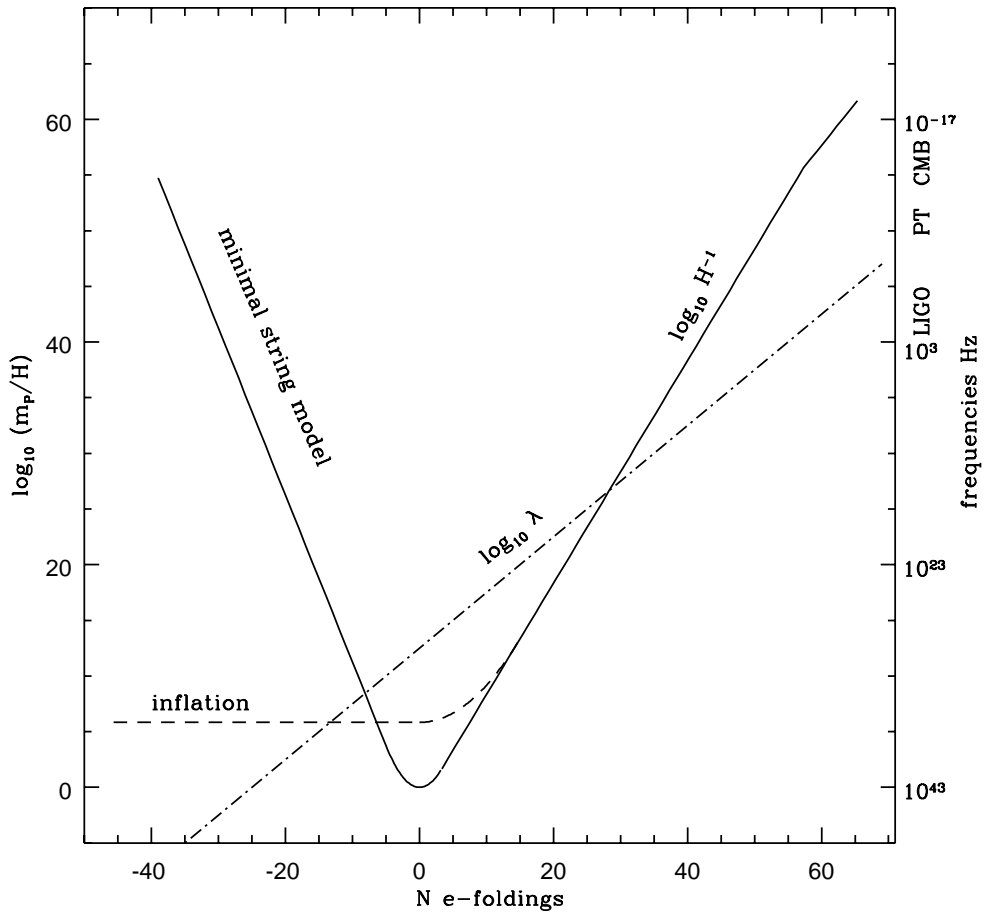}
\caption{\baselineskip 15.pt
The behavior of the Hubble horizon and of the 
physical wavelength in the cases of standard inflation 
and  deflation followed by the canonical radiation dominated 
era.}\label{wave1}
\end{figure}

The largest 
value of $k$ depends on the value of $H$ at the end of (deflation) inflation.
Typical values in the range $(10^{-8}\div 1) m_P$, i.e. temperatures
$(10^{15}\div 10^{19})$Gev, imply frequencies in the
range $(10^{7}\div 10^{11})$Hz, Fig.\ref{wave1}. Thus it 
makes sense to consider the possibility
of a detection of the gravitational background by the LIGO detectors that
work in the range $(1\div 10^3)$ Hz \cite{Giaz}.
Such a detection could fix spectral parameters, thus defining 
the cosmological background.
Quadrupole anisotropies in the Cosmic Microwave Background (CMB) and Pulsar
Timings (PT) already constrain the ranges of the spectral parameters,
but the region  they allow  in the plane of the parameters is still
large \cite{bggv}.

\subsection{Spectrum from the stringy era}
As discussed in the Introduction, the stringy era is better studied in the EF.
Two relevant consequences follow from {\it ansatz} (\ref{eq1b}).

\begin{figure}
\hskip 2truecm
\epsfxsize=11cm
\epsffile{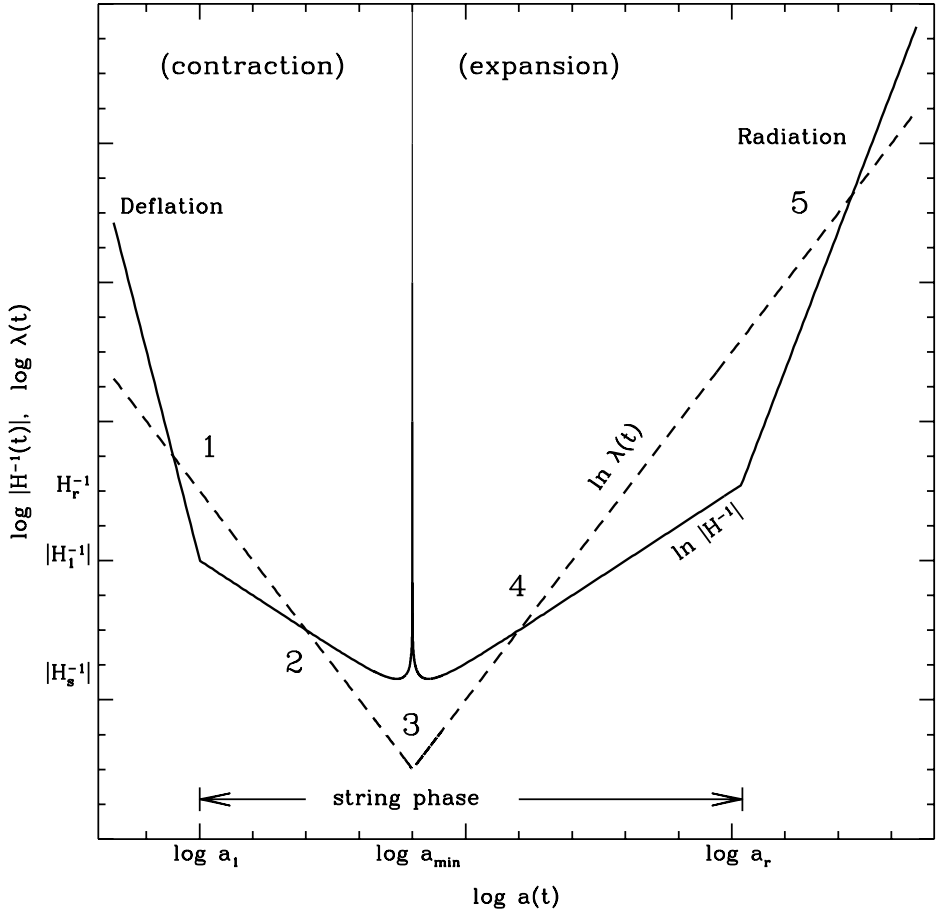}
\caption{\baselineskip 15.pt
A typical gravitational wave generated
during the PBB deflationary presents the features described in the 
test.}\label{wave}
\end{figure}

First, $H^{-1}$ goes to zero, as $t\to 0$, slower than the physical wavelength
$\lambda=a (2\pi/k)$: 
\begin{equation}\label{eq2}
\frac{\lambda}{H^{-1}}\sim \frac{4\pi}{t_0k}\sqrt{a-a_{min}}\,\,.
\end{equation}
This means, Fig.~\ref{wave}, that any wave, that crossed out the horizon in the 
deflationary stage at $(1)$, reenters it at $(2)$, since for $t<0$ 
$a\to a_{min}$ at $(3)$. 

On the contrary, for $t>0$, $a$ grows and alongwith the perturbation 
reexits at $(4)$. 
The subsequent evolution matches the ordinary picture, with a final 
{\em hallo again} in the radiation (or matter, depending on $k$) dominated era 
at $(5)$.
Something similar happens in the  case of double 
inflation \cite{zm} where the
subhorizon crossing regards a certain range of wavenumbers, and produces a
typical modulation in the spectrum. Here, due to deflation,
all the waves cross the subhorizon region. 

Secondly, according to (\ref{eq1b}),
there is always an era of power law inflation after a deflationary 
period, when $t>0$, since $a\sim t^2$. Thus the correct spectrum has the form 
of a deflationary one at small $k$ and of a power-law inflationary one at
large $k$. 
We underline that,
in order to obtain the correct description of the physics involved, we  must 
take into account both the deflationary and the subsequent power--law
inflationary era, whereas by neglecting the latter we would miss the signature 
in the spectrum described in this paper. This signature consists of a peak 
in the spectrum at a frequency $\omega_M$.


For very small frequencies $(\omega\ll 10^{-16}$, waves reentering during the 
matter dominated era)  the slope of the spectrum  
is easily obtained from (\ref{eq1})
with $b=1/3$, and $c=2/3$ 
$\Omega_g(\omega)\propto \omega$, while for frequencies in the range 
$\omega \in\{10^{-16}, \omega_M\}$ we have $b=1/3$, $c=1/2$  and then 
$\Omega_g(\omega)\propto \omega^3$. 
Similar results were found in a
different context \cite{bggv}. There,
the maximum frequency (one graviton created per Hubble volume) 
is given, Fig.~\ref{spettro1}, by 
\[
\omega_r=10^{11}\sqrt{\frac{H_r}{m_P}}\qquad\mbox{Hz}\,\,,
\]
where the subscript $r$ refers to the beginning of the radiation dominated 
era, and $m_P$ is the Planck mass: furthermore  $\omega_r $ coincides 
with the frequency of the peak.
In our model, the 
frequency of the peak is related to the free parameters by 
\[
\omega_M=\frac{10^{11}}{({z_{st}z_{infl}})^{1/2}}\sqrt{\frac{H_r}{m_P}}\qquad\mbox{Hz}
\,\,,
\]
where
\[
1+z_{st}=\left(\frac{a_{stend}}{a_{min}}\right),\quad 1+z_{infl}=
\left(\frac{a_r}{a_{stend}}\right)\,\,
\]
are the two parameters of the model and 
$a_{stend}$ and $a_{min}$ are respectively the value of the scale factor 
at the end of pure stringy era
(beginning of the power--law string era), and  
its minimum value at $t=0$, and $a_r$ is the value of the scale factor at the 
beginning of the radiation era.
The value of spectral density at the peak is:
\[
\Omega_M=\Omega_\gamma\left(\frac{H_r}{m_P}\right)^2{z_{st}^{7/2}z_{infl}}
\]
where $\Omega_\gamma h_{50}^2\simeq 10^{-4}$ is the present ratio of the 
photon density to critical density.

For the waves crossing the horizon during the string era we must consider 
the growing of amplitudes due to contracting metric [$a(t)\sim (-t)^2$] for 
wavelength larger than the Hubble horizon. The slope of the spectrum can be 
derived from (\ref{neweq}) with $g=2$, $\beta=3$ and results to be   
$\Omega_g(\omega)\propto \omega^{-7}$. 

Finally for very large frequencies the power--law inflation 
determines the slope 
$\Omega_g(\omega)\propto \omega^{-2}$, [from (\ref{eq1}) with $b=2$, 
$c=1/2$].
Summarizing:
\begin{eqnarray}
 \omega_o<\omega<\omega_e&\quad\quad&\Omega_g\simeq\Omega_\gamma\left(
  \frac{H_{ r}}{m_P}\right)^2
  \left(\frac{\omega_e}{\omega_{\mbox{\scriptsize M}}}\right)^3
  \left(\frac{\omega}{\omega_e}\right)
  {z_{st}^{7/2}z_{infl}}, \nonumber\\
 \omega_e<\omega<\omega_{\mbox{\scriptsize 
M}}&\quad\quad&\Omega_g\simeq\Omega_{\gamma}
  \left(\frac{H_{ r}}{m_P}\right)^2
\left(\frac{\omega}{\omega_{\mbox{\scriptsize M}}}\right)^3 
{z_{st}^{7/2}z_{infl}}, 
\nonumber\\
 \omega_{\mbox{\scriptsize M}}<\omega<\omega_1&\quad\quad&\Omega_g\simeq\Omega_{\gamma}
  \left(\frac{H_r}{m_P}\right)^2\left(\frac{\omega}{\omega_M}\right)^{-7}
  {z_{st}^{7/2}z_{infl}},\nonumber\\      
 \omega_1<\omega<\omega_r&\quad\quad&\Omega_g\simeq\Omega_{\gamma}
  \left(\frac{H_r}{m_P}\right)^2
  \left(\frac{\omega}{\omega_M}\right)^{-2}({z_{st}z_{infl}})^{1/2},
\end{eqnarray}
where $\omega_e=10^{-16}$Hz.

\begin{figure}
\hskip 2truecm
\epsfxsize=11cm
\epsffile{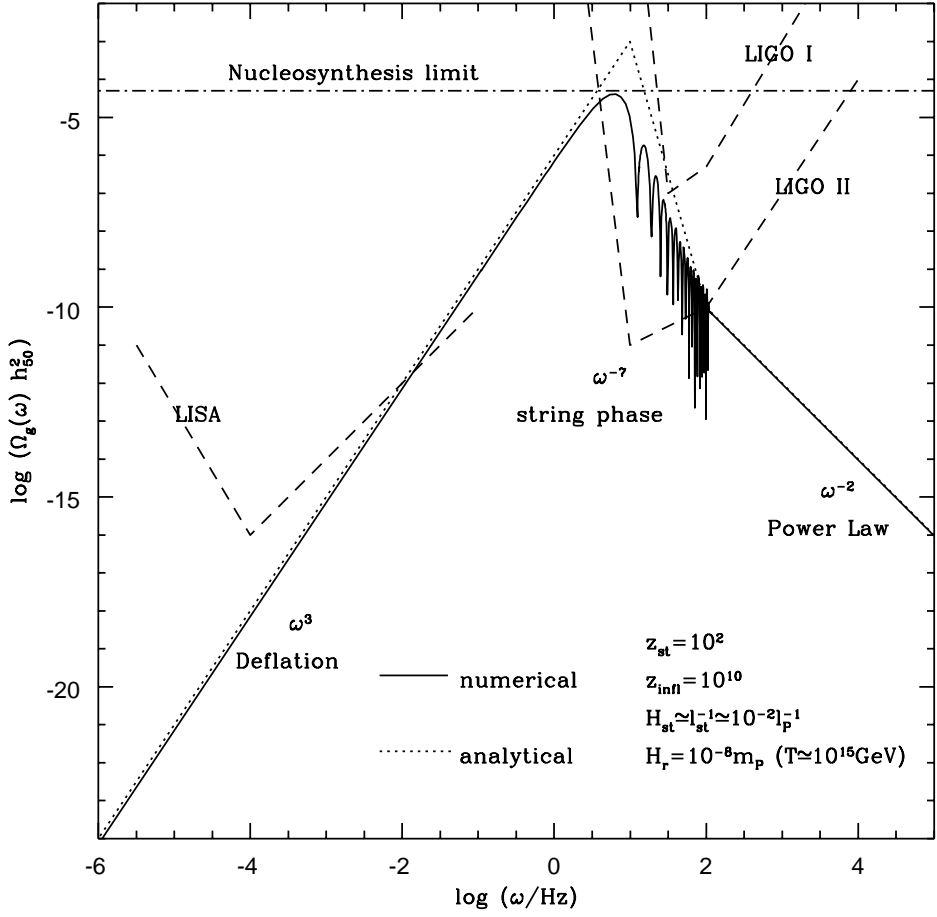}
\caption{\baselineskip 15.pt
The graviton spectrum of the stochastic background, with the
structured peak due to the stringy era,
superimposed to the known observational bounds and to the expected 
sensitivities of the LIGO's.}\label{spettro}
\end{figure}

The full spectrum is shown in Fig.\ref{spettro}. Three features 
emerge: 1) the growth due to deflation at small $\omega$;
 2) the existence of a sharp peak at $\omega_{M}$;
 3)  the descent due to  power-law inflation at large $\omega$.
 Concerning 2), we underline that the maximum of the spectrum 
$\Omega(\omega_{M})$ is higher than expected. This is due to the fact
that the perturbations that exit the horizon in the
stringy era are amplified more than those coming from the deflationary era.
 Furthermore, a comment must be made about the oscillations showing up in 
 Fig.~\ref{spettro}: they arise because
waves of different $\omega$  and different phases  
remain subhorizon for different time lengths. Finally,
the total, deflation+power-law, duration $N_T$ must be at least of 60 e-folds.

\begin{figure}
\hskip 2truecm
\epsfxsize=11cm
\epsffile{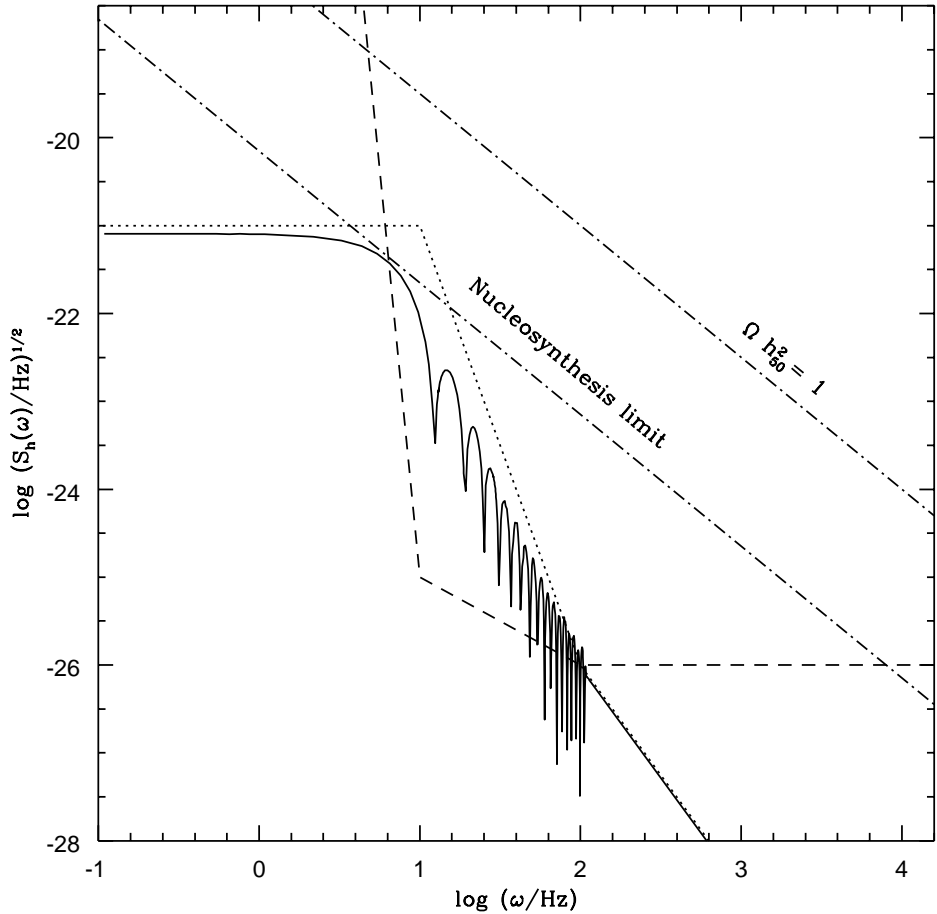}
\caption{\baselineskip 15.pt
The spectral amplitude of the stochastic background of Fig. 4}
\end{figure}

\begin{figure}
\hskip 2truecm
\epsfxsize=11cm
\epsffile{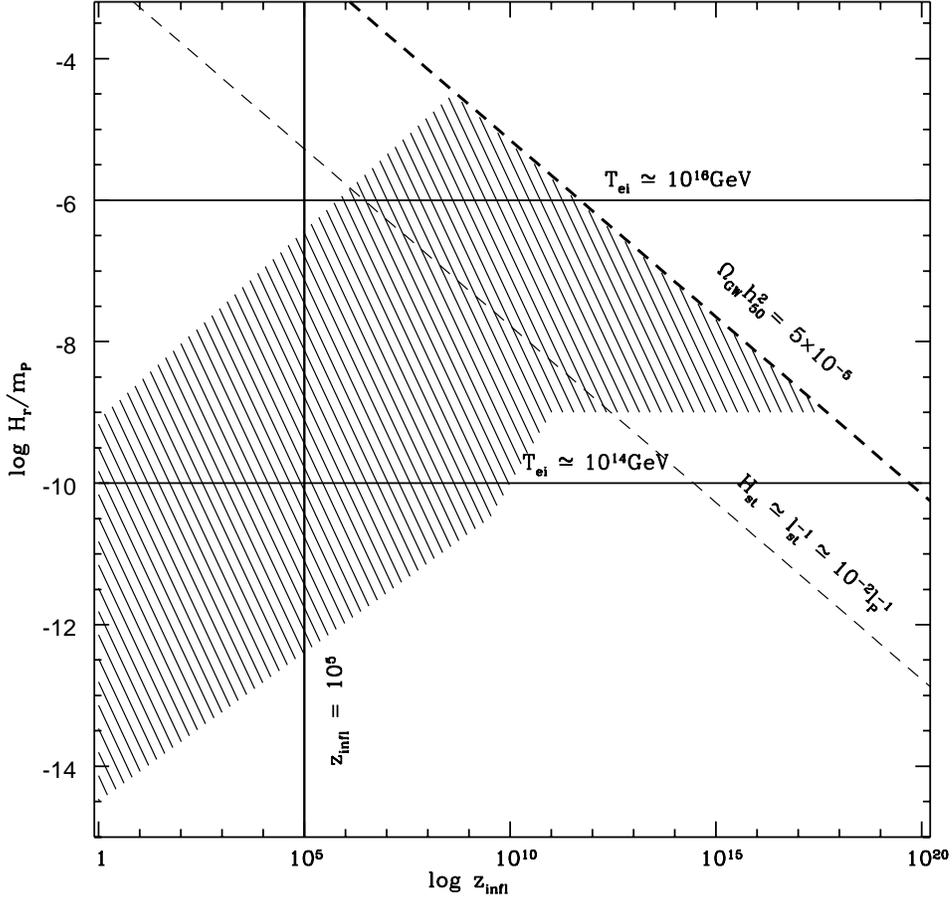}
\caption{\baselineskip 15.pt
The shading shows the parameter region for which the structured
peak could be detected by LIGO. 
A sensible model falls in the area 
$z_{infl}>10^5$ (monopoles), $10^{-10}<H_r/m_P<10^{-6}$ (GUT), and 
$\Omega_gh_{50}^2<5\cdot 10^{-5}$ (nucleosynthesis). 
The final constraint $10^{-2}<g_s<10^{-1}$ gives
the region between the dashed lines.}\label{param}
\end{figure}

This is a three parameter model: $N_{st}=\ln z_{st}$, the number of e-folds 
of the stringy era, $N_{infl}=\ln z_{infl}$, the number of e-folds of 
the power-law era; $H_r$, the value of the Hubble parameter at the end of 
inflation and the starting of the radiation dominated era. 
The shaded area in the Fig.~\ref{param} shows the region of the parameter 
space ($z_{infl},H_r$)
for which the peak of the spectrum falls in the range of detectability 
of LIGO/VIRGO interferometers \cite{allen}, for all $z_{st}$ values.

In addition, we must require that the graviton energy density does not exceed
the nucleosynthesis bound \cite{contr}, ($\Omega_{g}h_{50}^2< 5\times 10^{-5}$ 
and that the
inflationary era solves the monopole problem (limit from the nucleosynthesis,
$N_{infl}\geq 10$, i.e. $z_{infl}\geq 5$). Furthermore the low energy limit of
string theory and GUT models constrain $H_r$ to be in the range
$10^{-10}<H_r/m_P<10^{-6}$, that means $10^{-16}<T/GeV<10^{-14}$ \cite{chal}.
Finally, the value of $z_{st}$ is such that the maximum of $H=H_{st}\simeq l_{st}^{-1}$
is comparable to the string value $10^{-2}<H_{st}/m_P<10^{-1}$. 
The region defined by the above constraints has non zero intersection with the
shaded region. 
 
\section{Conclusions}
We have shown that the full stringy phase ($t\simeq 0$) is fundamental for 
detecting a primordial signature of string cosmology, while it is commonly 
assumed that only the PBB, pure dilaton era is relevant. 

To obtain this result
 we have studied both numerically and analytically the
spectra of primordial gravitons and we have proved that the latter has a
distinct peak. Our model has three free parameters: we find the region of
parameter space which satisfies all the known constraints. In this region both
the frequency and the amplitude of the peak may fall within the realm of the
new interferometric detectors. A typical value for the peak is
$h^2_{50}\Omega_M\simeq 5\times 10^{-5}$ at the frequency $\omega_M\simeq 10$
Hz. 

Finally we remark that, for simplicity's sake, we have 
framed our formulation
in a four dimensional space--time. In reality, our results hold true for any
$D\geq  4$ dimensional space--time which allows the compactification of the
internal dimensions \cite{prep}. This is so because, when the proper 
conformal frame is chosen
\cite{soko1},\cite{cho},  the gravitational wave equation still
takes the form (\ref{clwa}), 
that is independent of $D$.

\vspace{.5truecm} 
\centerline{\bf Acknowledgements}
We thank L. Amendola, V. Ferrari  and  E. W. Kolb  for enlightening 
discussions.

M.L. acknowledges support from DOE and NASA, grant NAG5-2788, at Fermilab.

\end{document}